# Effects of chemical disorder and spin-orbit coupling on electronic-structure and Fermi-surface topology of YbSb-based monopnictides

Maxwell Eibert,[1] Christopher Burgio,[1] Tyler Del Rose,[1] Prince Sharma,[1]
Yaroslav Mudryk,[1] and Prashant Singh[1,*]

[1]*Ames National Laboratory, U.S. Department of Energy, Iowa State University, Ames, IA 50011, USA*

**Abstract**

In this work, we study the influence of disorder on the electronic structure of YbSb—a rare-earth monopnictide featuring a simple rock-salt (B1) crystal structure and a well-defined Fermi surface topology—by employing first-principles density functional theory (DFT). We focus on chemical disorder introduced through Te and Al doping, selected based on their thermodynamic stability in alloyed configurations, to understand how such perturbations modify the electronic states of YbSb. Our results indicate that Te doping predominantly introduces electron-like states at the X and L points, while Al doping leads to a suppression of hole-like states at Γ, effectively driving the system from a semimetallic state to one characterized by very narrow-gap behavior at Γ. This modulation of the Fermi surface, particularly the reduction of central hole pockets at Γ, plays a central role in altering inter-pocket scattering—a mechanism critical for tuning quantum transport properties, including superconductivity. This disorder-driven modulation of the Fermi surface, particularly the suppression of central hole pockets at Γ controls inter-pocket scattering, which is essential for optimizing quantum transport properties, including superconductivity. Our results show that disorder can be effectively used as engineering band topology, thereby, tuning quantum related response through tailored electronic structure.

# 1. Introduction

Rare-earth monopnictides (LnPn, where Ln = Ce, Pr, Sm, Gd, Yb and Pn = As, Sb, Bi) have been extensively studied due to their diverse electronic and magnetic properties [1-8]. These materials exhibit phenomena such as large magnetoresistance, topological states, and Weyl fermion behavior, making them compelling candidates for quantum materials research [9-24]. The interplay between conduction electrons and localized *f*-electrons in LnPn compounds gives rise to complex behaviors, including exotic surface states and topological phases, as observed in LaBi and CeSb [25-28]. Furthermore, asymmetric mass acquisition in LaBi [29] and the presence of node-line states [30] highlight the intricate electronic structure of these materials.

Topological phase diagrams of rare-earth monopnictides emphasize the need to understand how electronic topology evolves across different compositions [31]. The extreme magnetoresistance (EMR) observed in LaSb [32] and the nearly perfect carrier compensation in LuBi and YBi semimetals [33] suggest that charge compensation plays a crucial role in their transport properties [23]. Interestingly, EMR is not exclusive to topological materials, as evidenced by LaAs, a topologically trivial system exhibiting similar behavior [22]. Studies on atomic orbital contributions near the Fermi level further refine our understanding of electronic correlations in these compounds [24]. Magnetic interactions also significantly influence their electronic characteristics, as seen in ferromagnetic DySb and NdSb, where Shubnikov-de Haas oscillations and Dirac-like semi-metallic phases emerge [34-36].

From a band structure standpoint, density-functional theory (DFT) calculations and ARPES have confirmed the semimetallic nature of many LnPn systems, typically featuring electron pockets at X and hole pockets at $\Gamma$ or L, with energy offsets depending on the specific compound [37-39]. While the effects of spin-orbit coupling (SOC) and magnetic ordering have been explored in these materials, however, the impact of chemical disorder—particularly site-selective substitutions—on the electronic structure and Fermi surface topology remains less understood. Controlled doping presents a promising route to modulate carrier concentration, alter phase stability and band topology, potentially inducing exotic states such as topological insulators and Weyl semimetals. This tunability opens new avenues for tailoring quantum materials for specific applications.

In this study, we focus on YbSb, a heavy rare-earth monopnictide crystallizing in a cubic rock-salt (B1) structure, characterized by a simple Brillouin zone and a well-defined Fermi surface. We employ DFT to investigate how chemical disorder, introduced via Te and Al doping at the Sb site, affects its electronic structure and Fermi-surface topology. which is essential for understanding inter-pocket scattering and quantum transport phenomena. We introduce aluminum (Al) and tellurium (Te) as dopants based on their thermodynamic stability in YbSb and their known influence on electronic band structures in similar monopnictides. Our analysis, guided by thermodynamic stability analysis, explores phase stability and electronic structure modifications, providing insights into how disorder-driven band topology engineering can enhance functional properties in rare-earth monopnictides.

2. **Methods**

*Computational method*: We employ DFT as implemented within Vienna Ab initio Simulation Package (VASP) for electronic-structure calculations (both with and without SOC), where the valence interaction among electrons were described by a projector augmented-wave method [40,41] with a plane-waves energy cutoff of 520 eV. For full relaxation of compounds, we set very strict convergence criteria of total energy and force convergence, i.e., $10^{-8}$ eV/cell and $10^{-6}$ eV/Å. In (semi)local functionals, such as GGA, the *f*-electrons are always delocalized due to their large self-interaction error [42,43], so we employ the Perdew-Burke-Ernzerhof (PBESol) exchange-correlation functional in the generalized gradient approximation (GGA) [44]. For clarity, we want to emphasize the use of Yb_2 pseudopotentials as the findings align very well with DFT [45] and others [46] as well as with the experiments reported by Abulkhaev [47] and Leon-Escamilla et al. [48]. To enforce the localization of the *f*-electrons, we perform PBE+U calculations [with a Hubbard U (Yb=X eV, Sb= eV; J=0.9 eV) introduced in a screened Hartree-Fock manner [49,50]. We used $11 \times 11 \times 11$ ($7 \times 7 \times 7$) Monkhorst-Pack k-mesh for Brillouin zone sampling [51] of Yb-Sb (Yb-Se-Te-Al) compounds during self-consistent electronic calculation while $7 \times 7 \times 7$ ($5 \times 5 \times 5$) k-mesh was used for full (volume and atomic) structural relaxation. The partially ordered structures of Yb($Sb_xTe_yAl_{1-x-y}$) for electronic-structure calculations was site-substituted using Alloy Theoretic Automated Toolkit (ATAT) [52]. We used Sumo [53] and Fermisurfer [54] plotting programs to postprocess the electronic-structure and Fermi surface diagrams.

*Experimental method*: To confirm the phase stability of YbSb compound, experimental synthesis was attempted using two different routes. The first method involved melting the components in a tube furnace and resulted in primary formation of the $Yb_{11}Sb_{10}$ phase. In second method chunks of Yb and Sb taken in stoichiometric amounts were cut into small pieces and ball-milled using a silicon nitride vial to avoid magnetic contamination. Next, the milled powder was pressed using hand press in Ar-filled glovebox to ensure maximum surface contact of ingredients during the reaction process. The pellet was taken out and sealed in quartz ampoule under partial Ar atmosphere then placed in the same muffle furnace at 700° C for 72 hours to allow a reaction of molten Sb with milled Yb to occur. Next, the sample was opened in a glovebox and mortared into a fine powder using agate mortar and pestle to prepare it for X-ray powder diffraction (XPD) analysis. X'Pert Pro PANalytical diffractometer using Cu Kα radiation was employed to collect the data using an air sensitive sample holder. The XPD patterns obtained were analyzed using JADE analysis software and Rietveld refinement was performed using FullProf.

## 3. Results and discussion

The face-centered cubic rock-salt (B1) crystal structure is commonly adopted by rare-earth monopnictides (LnPn, where Ln = La, Yb, etc., and Pn = Sb, Bi, etc.) (**Fig. 1a**). The Ln and Pn atoms form interpenetrating cubic sublattices, leading to a highly symmetric and dense atomic packing. The stability of the rock-salt phase is governed by factors such as ionic size, electronegativity differences, and electronic hybridization between Ln-*d or f* states and Pn-*p* states. In this study, we specifically focus on YbSb as a representative system within this class and experimentally validate its structural phase through powder X-ray diffraction (XRD) and magnetic characterization as shown in **Fig. 1**. The XRD pattern of YbSb in **Fig. 1b**, analyzed via Rietveld refinement, shows excellent agreement between the observed (red circles) and calculated (blue line) profiles. The minimal residuals (green line) confirm the formation of a single-phase, highly crystalline YbSb compound. The diffraction peaks are sharp, well-indexed, and consistent with the expected cubic symmetry, with no evidence of secondary phases or impurity-related reflections, thereby affirming the phase purity and structural integrity of the synthesized YbSb.

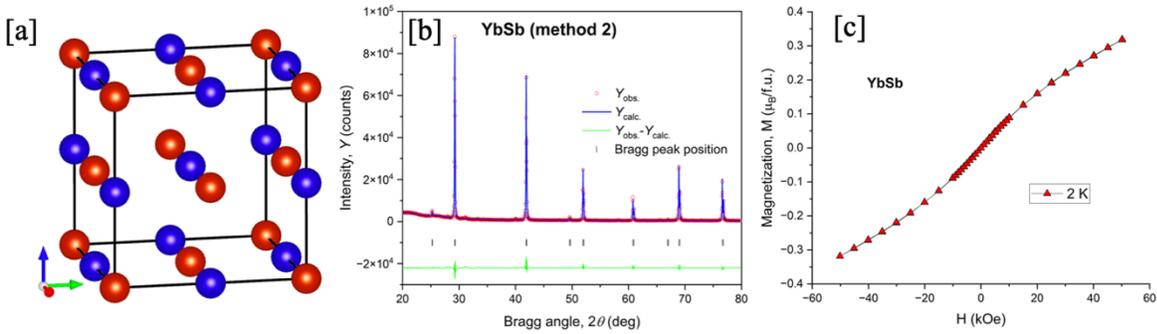

**Figure 1.** (a) Rock-salt crystal structure of YbSb. (b) Rietveld-refined XRD pattern confirming phase-pure, cubic YbSb. (c) Magnetization vs. field at 2 K showing correlated paramagnetic, i.e., weak magnetic response, is consistent with DFT.

In **Figure 1c**, we show the field-dependent magnetization of YbSb at 2 K. The magnetization curve is smooth and nonlinear, reaching approximately 0.3 $\mu_B$ per Yb ion at ±50 kOe, without any signs of hysteresis. This behavior is characteristic of a strongly correlated paramagnetic state rather than a conventional ferromagnet. The non-saturating response and lack of remanence suggest the presence of strong spin correlations or fluctuations, potentially tied to Kondo screening or antiferroquadrupolar interactions, which are consistent with the known low-temperature properties of YbSb. This experimental trend is in good agreement with our DFT-based magnetic moment calculations, which yields a Yb magnetic moment of approximately 0.23 $\mu_B$, further supporting a partially compensated 4*f* moment. In SI **Fig. S5**, we show magnetic measurements of the YbSb as a function of temperature in several applied magnetic fields, revealing weak magnetic response. . The non-linear nature of $\chi^{-1}$ vs. T further suggest the Yb sublattice support weak magnetic interactions at low temperatures. This weak magnetic nature of the $Yb^{3+}$ ions with no clear magnetic ordering is in agreement with literature [54, 55]. The phase pure rock salt structure in combination with the confirmation of weak magnetic nature of the Yb atoms line up well with the results published in literature, indicating that a direct solid-state reaction is a viable alternative to the standard melt reactions in prepare YbSb samples. In fact, given the volatility of the constituent elements, it is expected that this solid-state reaction is to be the preferred synthesis route for future YbSb and YbSb-based compounds.

Following the structural and magnetic characterization of YbSb in **Fig. 1**, which confirmed the formation of a phase-pure cubic rock-salt lattice and revealed correlated paramagnetic behavior at low temperatures, we turn to first-principles calculations to investigate the underlying electronic structure. **Figure 2a** shows the DFT-calculated band structure of YbSb, highlighting the role of spin–orbit coupling (SOC) in shaping the low-energy states and Fermi surface features. The first Brillouin zone is constructed from primitive cell of the reciprocal lattice (see inset **Fig. 2a**), using the Fourier transform of the crystal lattice. The dimensions of the BZ are the reciprocal of the dimensions of the crystal lattice in **Fig. 2a**, i.e., sides are the dimension "$1/a_{lat}$". The Greek and Roman letters on the edge of the BZ are the high-symmetry Wycoff points, which are useful in determining electronic-structure and Fermi-surface topology of band structure.

In **Fig. 2a** (no-SOC), the band structure is plotted along high-symmetry directions (Γ-K-L-U|W-W2-X) and exhibits multiple band crossings, particularly near the Fermi level, as highlighted by the blue boxes. Several bands meet at the same energy and momentum, forming degeneracies that suggest possible Dirac-like behavior, especially near U and W2. The conduction band minimum (CBM) and valence band maximum (VBM) appear relatively close, indicating a small bandgap or a semimetallic nature. However, experimental studies reveal an unexpected splitting of the $Γ_8$ crystal field states, suggesting that additional mechanisms may be involved [57]. Notably, SOC included band-structure in **Fig. 2b** shows that some nearly degenerate bands in the no-SOC case now show fine splitting's on the order of hundreds of meV, a characteristic feature of spin-orbit interactions. The SOC-induced band splitting for Yb-Sb in **Fig. 2b** is significant, with quantitative splits ranging from approximately 50 meV to 150 meV near the Fermi level, as clearly demonstrated by the change from nearly degenerate bands in the absence of SOC (**Fig. 2a**) to well-separated branches in its presence (**Fig. 2b**). These splitting, arising from the strong intrinsic spin-orbit interaction in Yb's heavy 4*f* states, lift degeneracies at high-symmetry *k*-points such as K, along L-U/W, and near X, ultimately altering the Fermi-surface details and impacting low-temperature properties such as effective masses and hybridization.

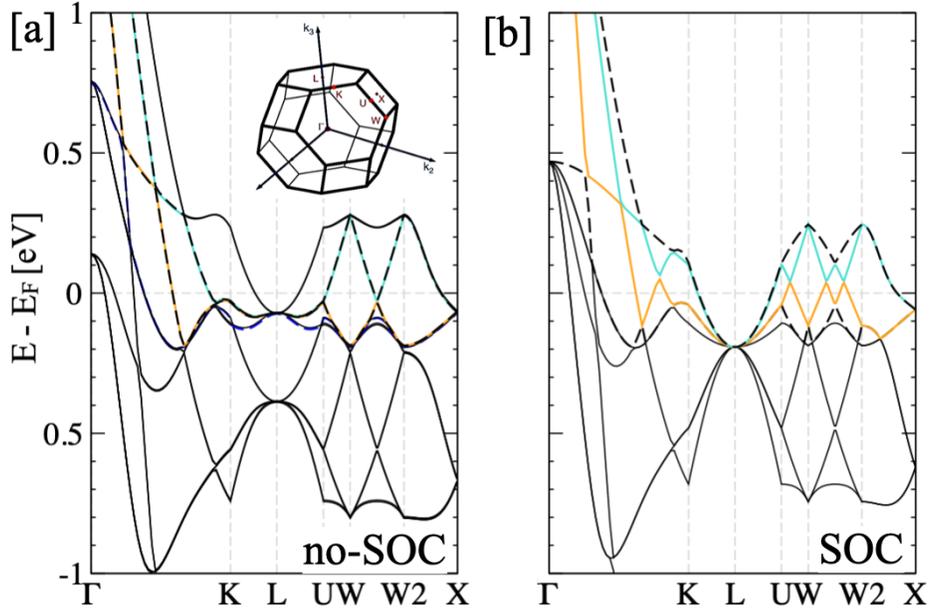

**Figure 2**. DFT-calculated band structure of YbSb (a) without and (b) with spin–orbit coupling (SOC) along high-symmetry paths (Brillouin zone inset). SOC lifts degeneracies and induces band splittings up to ~150 meV near X, L, and W points, significantly altering the low-energy electronic structure and Fermi surface topology.

While dispersive effects in the excitation spectrum may partially account for this behavior, the possibility of additional symmetry-breaking mechanisms, such as higher-rank multipolar orders or subtle lattice distortions, could further split or shift the $\Gamma_8$ levels. Direct measurements using high-resolution inelastic neutron scattering or Raman spectroscopy are needed to resolve whether these effects arise solely from SOC or if additional hidden orders contribute, thereby providing a more complete understanding of the low-temperature phase behavior in Yb–Sb.

To further understand the role of SOC on electronic-structure of YbSb, we focused on energy bands near the Fermi-level in **Figure 3**, which provides a detailed analysis of the band structure, charge density, and Fermi surface, complementing the discussion of SOC-induced modifications in electronic structures in **Fig. 2**. Notably, the band structure in **Fig. 3a** along high-symmetry points (K, L, U, W, W2, X) exhibits spin-orbit interaction effects, including band splitting and shift relative to the Fermi level. Compared to the no-SOC case, degeneracies present in previous discussions are now lifted, with noticeable changes near the W and W2 points. The conduction and valence bands show modifications in dispersion, which impact carrier dynamics and potential topological properties. Notably, the zoomed-in region in **Fig. 3b** focuses on band structure near the W point, highlighting a linear band crossing that is indicative of Dirac-like or Weyl-like behavior. The inset provides a magnified view of this feature, showing SOC-driven band splitting and possible band inversion. This supports the hypothesis that SOC can induce nontrivial topological phases by modifying band connectivity and lifting degeneracies. The presence of a band crossing in this region suggests that this material may exhibit topologically protected states or exotic transport properties.

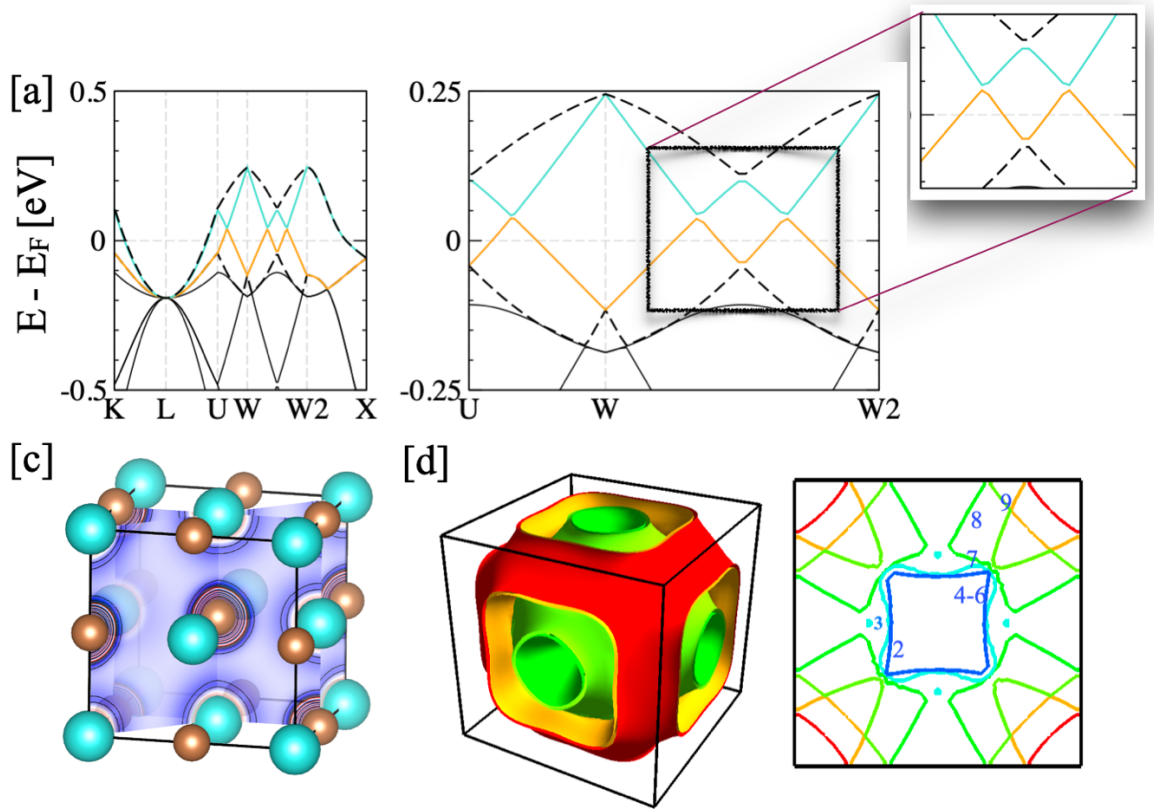

**Figure 3.** Spin-orbit coupled electronic structure analysis showing (a) the band structure along high-symmetry points, (b) a zoomed-in view near the W point highlighting SOC-induced band splitting, (c) charge density distribution revealing electronic localization, and (d) the Fermi surface with intricate topology, indicating possible topological semimetallic behavior (see SI **Fig. S1-S3**).

**Figure 3c** shows the charge density distribution in real space, illustrating how electronic states are distributed around atomic positions. The charge localization around specific atomic sites indicates hybridization effects between orbitals, which are crucial for understanding bonding characteristics and the influence of SOC on electronic properties. The interplay between charge distribution and band structure changes further supports the role of SOC in shaping the material's electronic behavior. Additionally, the SOC Fermi surface in **Fig. 3d** shows multiple distinct pockets, i.e., a complex topology, suggesting strong SOC-induced modifications. 3D rendering of the Fermi surface reveals anisotropic features, with possible electron and hole pockets contributing to charge transport. The projected Fermi surface contour map on the right further clarifies the connectivity of bands at the Fermi level, where different colors correspond to different energy contours. The presence of multiple Fermi pockets suggests a compensated semimetallic nature, where electron and hole carriers are nearly balanced. Given the previously discussed SOC-driven band splitting and possible topological features, the intricate structure of the Fermi surface may contribute to unconventional transport phenomena such as extreme magnetoresistance or topologically nontrivial behavior.

DFT calculated band structure near the Fermi level for YbSb in **Fig. 2b and Fig. 3a,b** exhibits SOC-induced splittings in the range of 50-150 meV, which are comparable to those in established Dirac semimetals such as $Na_3Bi$ or $Cd_3As_2$ [**59,59**]. In YbSb, strong SOC lifts degeneracies at high-symmetry k-points, and when combined with symmetry breaking from quadrupolar or higher-rank multipolar ordering (evidenced by unexpected splitting of the $\Gamma_8$ states at temperatures up to 10 K) [**60**], the conditions for protecting Dirac crossings can be lost. This loss may either open a gap or split the Dirac points into pairs of Weyl nodes with finite momentum separation, suggesting that even subtle changes in crystal symmetry-potentially induced by temperature, magnetic field, or lattice distortions-could trigger a transition to a topologically nontrivial state. Such phenomena are analogous to those discussed in the literature on topological nodal-line semimetals [**61**] and Weyl physics in complex materials [**62,63**], providing a quantitative framework that links the SOC-driven band splitting and multipolar symmetry breaking in YbSb to emergent Dirac and Weyl behavior.

The Fermi surface topology of YbSb is analyzed more exhaustively along different crystallographic orientations, comparing the cases without SOC in **Fig. 4**(a, c, e, g) and with SOC in **Fig. 4**(b, d, f, h). We found that the inclusion of SOC introduces significant modifications to the Fermi surface contours and connectivity, impacting the electronic structure and carrier transport properties. Without SOC, the 3D Fermi surface in **Fig. 4a** exhibits highly anisotropic features with distinct electron pockets at high-symmetry points. The 2D projection along the (001) plane in **Fig. 4c** shows a well-defined, square-like central feature with additional nested bands at higher energies, indicative of a complex multi-band electronic structure. The 2D cross-section along the (111) plane in **Fig. 4e** presents a hexagonal pattern, with multiple concentric contours suggesting strong band interactions. The (011) cut in **Fig. 4g** displays an elongated Fermi surface with significant anisotropy, highlighting the directional dependence of the electronic states. While with SOC included the 3D Fermi surface in **Fig. 4b** becomes more isotropic, with a reduction in the complexity of the electron pockets. The (001) cross-section in **Fig. 4d** shows a significant redistribution of states, with a central feature that is more rounded compared to the non-SOC case, suggesting enhanced electron localization. The (111) projection in **Fig. 4f** demonstrates a collapse of the multiple nested structures seen in **Fig. 4e**, consolidating into a single dominant Fermi pocket. The (011) cross-section in **Fig. 4h** reflects a similar trend, where the previously elongated features transform into a more uniform, circular shape, indicating a shift in the effective mass and carrier mobility.

Interestingly, the introduction of SOC results in a smoothing of the Fermi surface contours, reducing the anisotropy and modifying the band structure near the Fermi level. This transition suggests that SOC plays a critical role in shaping the electronic properties of YbSb, particularly in modifying the density of states, carrier effective mass, and transport anisotropy. These effects are crucial for understanding the material's quantum transport behavior and its potential applications in spintronic and thermoelectric devices.

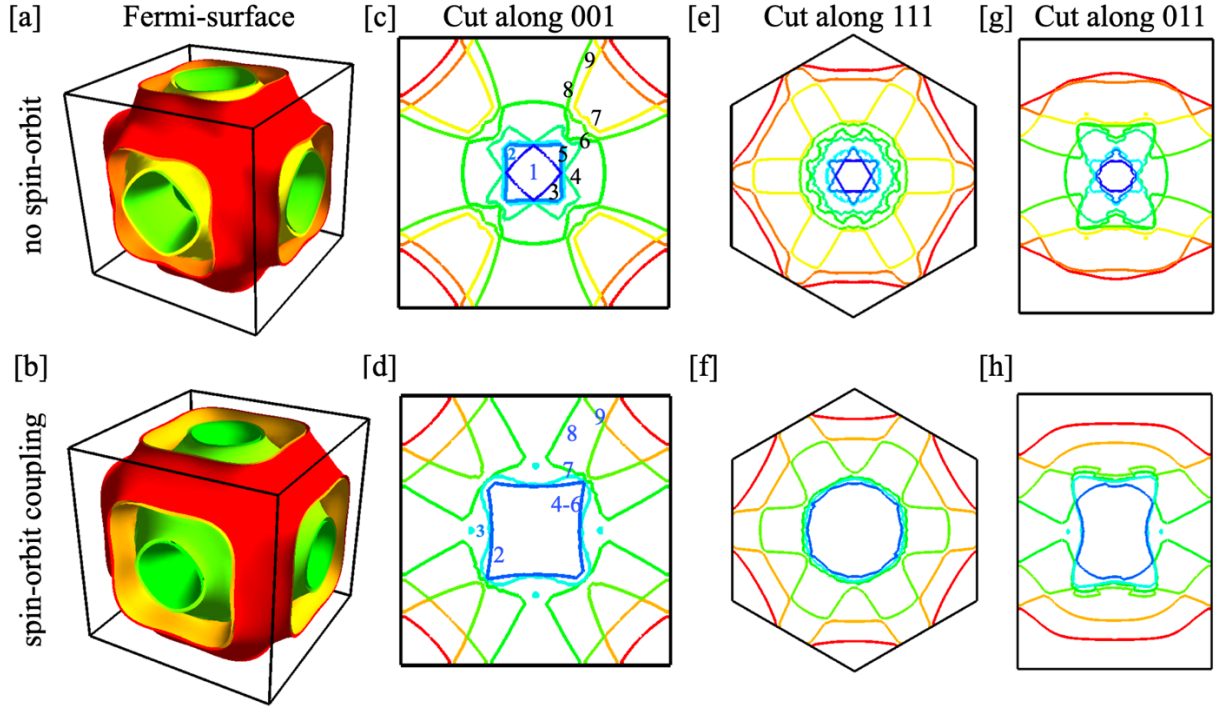

**Figure 4.** Comparison of the Fermi surface of without (a, c, e, g) and with (b, d, f, h) SOC along different crystallographic orientations. The SOC reduces anisotropy, smooths band contours, and modifies carrier dynamics, leading to a more isotropic electronic structure.

To understand the role of disorder, we choose YbSb as system of interest due to its simple crystal structure, and notable quantum behavior (band and Fermi-surface topology). One other reason for the choice of this system is typical reported disagreement between DFT and experiments [**45**]. **Figure 5** illustrates the band structures of YbX compounds (X = In, Sn, Sb, Te) with increasing valence electron count (VEC). The x-axis represents the progression from VEC = -2 (YbIn) to VEC = +1 (YbTe), showing how the electronic structure evolves. The Fermi level ($E_F$) is aligned in all four band structures, enabling a direct comparison. The X = In, Sn compounds crystallizing in B1 structure are hypothetical, while YbTe adopts the same rock-salt crystal structure as YbSb.

Starting with YbIn (VEC = -2), the band structure exhibits metallic behavior with multiple band crossings at the Fermi level. The conduction and valence bands are highly dispersive, and the absence of a bandgap confirms its metallic nature. Moving to YbSn (VEC = -1), the complexity of the band structure increases with additional band crossings near $E_F$. This suggests a transition toward a semimetallic state, where a very small or negligible bandgap may exist.

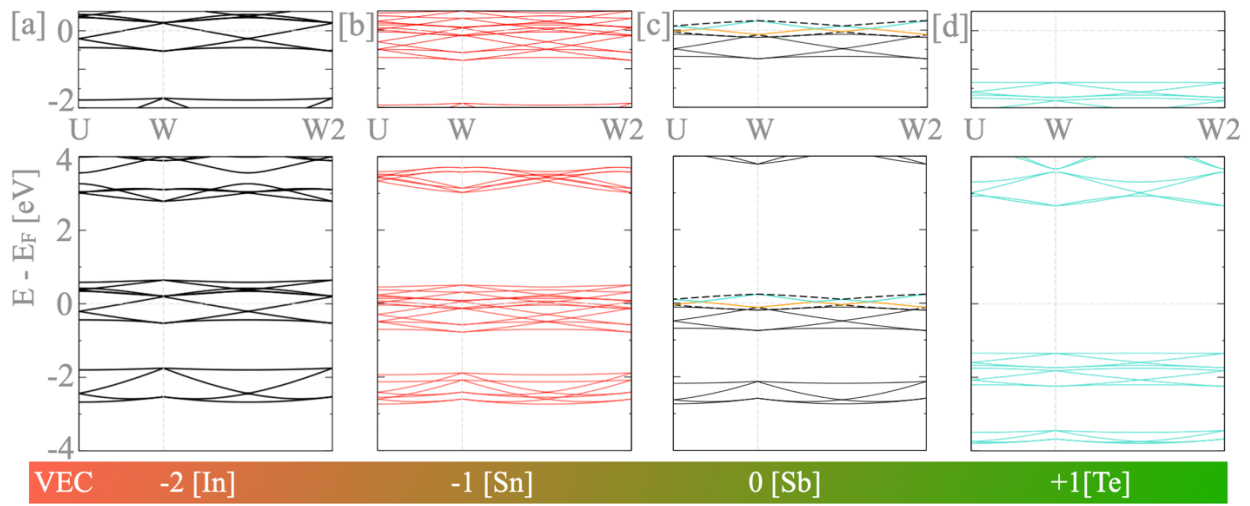

**Figure 5**. An SOC band-structure evolution of (a) YbIn, (b) YbSn, (c) YbSb, and (d) YbTe showing evolution (along U-W-W2 part of BZ) with increasing valence electron counts in YbX compound. The left of YbSb is hole-like doping while right to YbSb is electron-like doping, which is reflected in shift in bands around Fermi-level. YbX shows sharp change in band-structure with Te substitution.

In the case of YbSb (VEC = 0), a noticeable gap-like feature emerges, marking a shift toward a semiconducting phase. The presence of different colors in the bands may indicate orbital contributions or SOC effects. Although not fully developed, the bandgap is beginning to form, showing the material's intermediate electronic nature. Finally, YbTe (VEC = +1) exhibits a well-defined bandgap, indicating its transition to a semiconductor or insulator. The valence and conduction bands are clearly separated, and the overall band dispersion decreases, suggesting reduced electronic mobility. Quantitatively, the estimated bandgap for YbTe appears to be in the range of approximately 0.2–0.5 eV. As VEC increases, the number of band crossings systematically reduces, and the conduction and valence bands become flatter, which directly affects charge carrier mobility. Overall trend shows that increasing the electronic concentration of Yb-based compounds through substitution induces a transition from metallic (YbIn) to semimetallic (YbSn), then to a narrow-gap (YbSb), and finally to a wider-gap (YbTe) electronic-states at the Fermi-level in **Fig. 5**. This indicates that band engineering can be achieved through element specific substitution, making these materials tunable for desired electronic properties.

Thermodynamic analysis and convex hull assessment are essential for understanding phase stability, guiding material synthesis, and predicting possible design pathways [**64-70**]. **Figure 6a** shows the DFT-calculated convex hull for YbX (X = In, Sn, Sb, Te), showing the formation energy per atom ($E_{form}$). Among the four compositions, YbTe exhibits the most negative formation energy at approximately -1.8 eV/atom, indicating its high thermodynamic stability. YbSb is also stable but sits at a slightly higher energy compared to YbTe. In contrast, YbIn and YbSn have relatively less negative formation energies, suggesting they are less stable in the parent phase of YbSb (all thermodynamic calculations are performed on rock-salt crystal phase of YbSb). While **Fig. 6b** illustrates the thermodynamic stability of Yb(Sb$_{1-x}$Te$_x$) as a function of Te concentration. At a low Te content of around 3 atomic percent, the system shows a formation energy near -0.62 eV/atom. As Te content increases to approximately 6 atomic percent, the formation energy decreases further to around -0.67 eV/atom, implying that Te incorporation enhances stability within this composition range. This trend suggests that moderate Te substitution stabilizes the YbSb structure by lowering its overall formation energy.

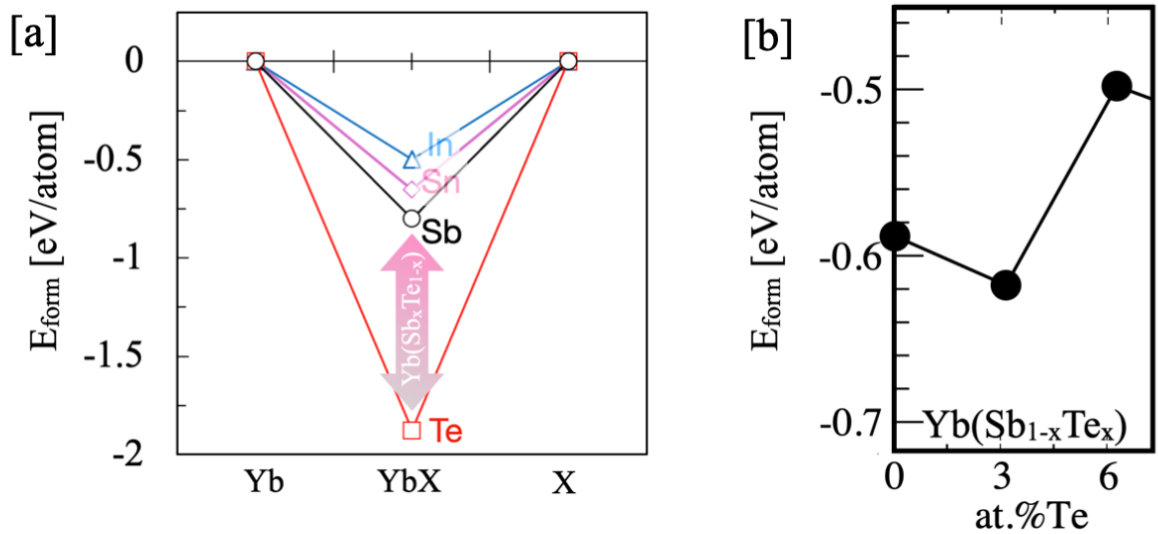

**Figure 6**. (a) DFT calculated convex hull of YbX (X=In, Sn, Sb, Te) compounds with respect to their endpoints illustrating the relative stability of different pnictogen and group-15/16 substituted phases. Te substitution shows the largest energy difference from Sb, indicating the strongest driving force for chemical perturbation. (b) Formation energy as a function of Te concentration in Yb(Sb$_{1-x}$Te$_x$), showing that up to ~3 at.% Te maintains thermodynamic stability, beyond which the compound becomes energetically less favorable.

The electronic band structures of YbSb, YbSb$_{0.93}$Te$_{0.03}$ and YbSb$_{0.93}$Te$_{0.035}$Al$_{0.035}$, **Fig. 7a-c**, exhibit significant modifications as disorder are introduced at the Sb site. The possibility of Weyl-like or topological features in YbSb and its disordered derivatives can be explored by drawing parallels to similar rare-earth monopnictides studied in literature. In materials such as LaBi, LaSb, and CeSb, nontrivial topology has been observed due to band inversion and SOC effects, leading to exotic electronic states such as Dirac-like dispersion and surface states [**16,31**].

The band structure of YbSb in **Fig. 7a** reveals significant modifications when SOC is introduced. Without SOC, the bands remain degenerate at the Γ-point and along high-symmetry directions in the Brillouin zone (e.g., M – Γ – R)., particularly near the Fermi level, indicating a semi-metallic nature. A Dirac-like crossing appears at Γ, suggesting the presence of a gapless state. When SOC is included, these degeneracies are lifted, leading to a notable splitting of approximately 0.15-0.25 eV at the Γ-point. The conduction and valence bands shift, with small energy gaps on the order of 20-50 meV emerging at other high-symmetry points too. The red-highlighted bands in the zoomed-in region suggest possible band inversion, a key feature of topological materials. The presence of such an inversion hint at a transition from a trivial semi-metal to a nontrivial topological phase. The electronic structure modifications due to SOC indicate that YbSb may exhibit nontrivial topology, while the introduction of disorder and chemical doping, as seen in **Fig. 7b&c**, could further influence the band structure and topological properties.

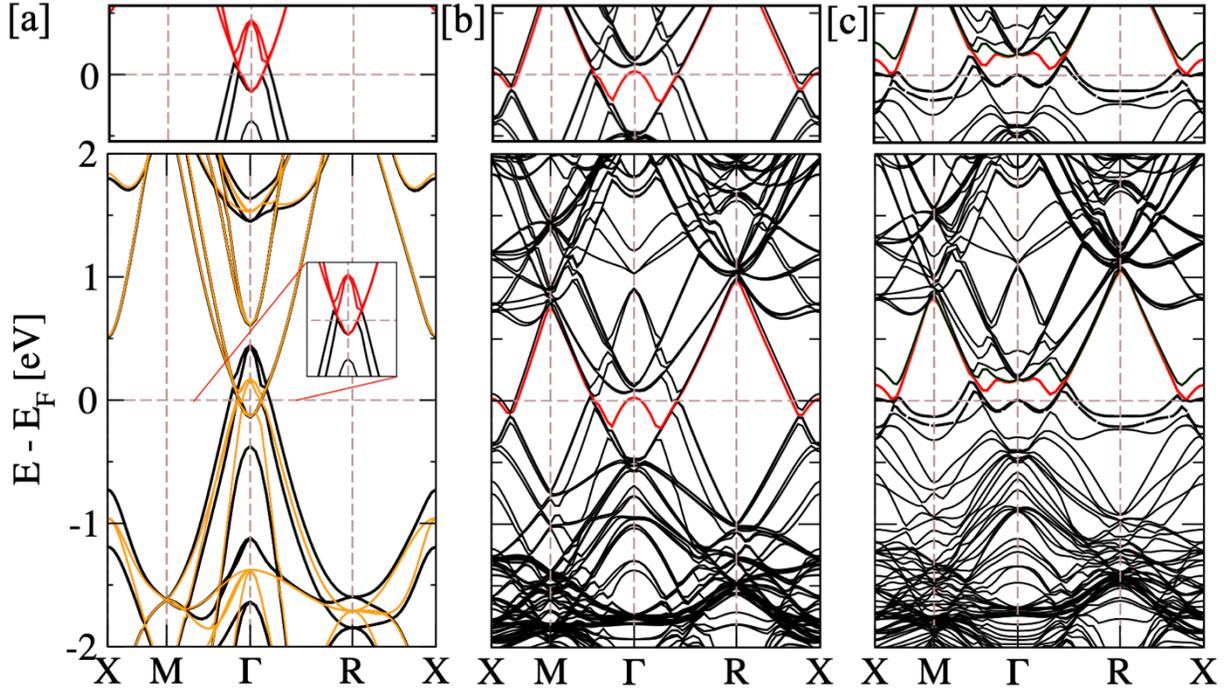

**Figure 7**. Electronic band-structure of (a) YbSb (orange line indicates calculations without SOC), (b) YbSb$_{0.93}$Te$_{0.03}$, and (c) YbSb$_{0.93}$Te$_{0.035}$Al$_{0.035}$ (both calculated with SOC), reveal disorder-induced modifications. The introduction of Te and Al leads to significant band broadening, suppression of hole-like states at Γ and X, and enhancement of electron-like states at M and R. These changes suggest a shift in Fermi surface topology, which, based on literature on rare-earth monopnictides, could indicate the emergence of Weyl-like features or topological electronic states.

Notably, Te substitution induces notable band broadening and increases band crossings, particularly near the M and R points, as shown in **Fig. 7b**. Similar effects in LaBi and LaSb have been linked to the emergence of nontrivial topology, as evidenced by angle-resolved photoemission spectroscopy (ARPES) and transport studies [**5,13**]. In disordered YbSb alloys, the altered band dispersion raises the possibility of topological surface states or Weyl-like features, potentially influencing Berry curvature and leading to anomalous transport effects such as chiral anomaly-induced negative magnetoresistance or anomalous Hall conductivity.

Compared to the SOC-included band structure of pristine YbSb (**Fig. 7a**), Te-substituted YbSb$_{0.93}$Te$_{0.03}$ (**Fig. 7b**) and co-doped and YbSb$_{0.93}$Te$_{0.035}$Al$_{0.035}$ (**Fig. 7c**) exhibit increased band broadening and subtle energy shifts near the Fermi level, particularly at the Γ-point. The red-highlighted bands indicate that while SOC-induced degeneracy lifting is retained, Te substitution alters band dispersion, likely due to local strain effects from charge redistribution. In **Fig. 7c**, corresponding to YbSb$_{0.93}$Te$_{0.035}$Al$_{0.035}$, disorder effects become even more pronounced. The conduction and valence bands appear more diffused, with additional band splitting and an increased bandgap at Γ. These modifications suggest significant shifts in the Fermi surface topology and the potential evolution toward a nontrivial electronic phase. In materials such as NdSb and DySb, similar band modifications have led to the emergence of Dirac-like states and magnetically tunable topological transitions [**16,17**]. While the fundamental SOC-driven band structure remains intact, Te and Al doping introduce perturbations that lift band degeneracies, modify electronic correlations, and influence both Fermi surface characteristics and the material's topological phase stability (see **Fig. 6**).

Finally, we examined the evolution of the Fermi surface in YbSb due to site-selective chemical disorder at the Sb site. The 3D (top panel) and 2D (bottom panel) projected Fermi surfaces of YbSb, Yb(Sb$_{0.97}$Te$_{0.03}$), and Yb(Sb$_{0.93}$Te$_{0.035}$Al$_{0.035}$) in the rock-salt crystal structure reveal significant modifications induced by Te and Al substitutions, as shown in **Fig. 8**. These changes suggest that disorder engineering plays a crucial role in tuning the electronic topology and transport properties of YbSb-based monopnictides.

**Figure 8a** illustrates the 3D Fermi surface of pristine YbSb, which features well-defined electron pockets centered around high-symmetry points, forming a connected topology characteristic of semimetals. The 2D projected Fermi surface along the Γ-X plane exhibits a symmetric pattern with sharp, well-defined contours. The presence of multiple Fermi pockets indicates complex carrier dynamics, arising from the interplay between light and heavy electron bands. With Te substitution (Yb(Sb$_{0.97}$Te$_{0.03}$), **Fig. 8b**), the Fermi surface becomes more fragmented, reducing connectivity between electron pockets. This transition suggests an increased isotropy in the electronic structure, leading to a reduction in the anisotropy observed in pristine YbSb. The 2D projection reveals a circular central feature, signifying band shifts that redistribute carriers. These changes suggest modifications in electronic transport behavior, which may reduce the effective mass and enhance carrier mobility. For the co-doped system Yb(Sb$_{0.93}$Te$_{0.035}$Al$_{0.035}$) (**Fig. 8c**), the Fermi surface undergoes substantial reorganization. The 3D structure exhibits increased distortions and additional features introduced by the combined effects of Te and Al substitutions. The 2D projection reveals new concentric Fermi pockets at the center, indicating enhanced band splitting and electron localization. These changes suggest a modified density of states at the Fermi level, which could influence electrical conductivity and topological properties. The progressive evolution of the Fermi surface topology—from a connected, anisotropic structure in YbSb to a fragmented, more isotropic topology in doped compositions—suggests substantial modifications in carrier dynamics. Te substitution disrupts Fermi surface connectivity, while additional Al doping further distorts the topology, likely introducing new scattering mechanisms and modifying band dispersion. Given the crucial role of inter-pocket scattering in quantum transport phenomena, these changes could significantly impact electrical resistivity, mobility, and superconducting behavior.

Notably, the intertwined and degenerate electron and hole pockets in YbSb, located at the center ($\Gamma$) and corner (X) of the Brillouin zone (BZ), may facilitate inter-pocket scattering [**71**], a key factor in high-$T_C$ superconductivity [**72,73**]. A similar mechanism is observed in Fe-based pnictides, where strong inter-pocket scattering between hole pockets at $\Gamma$ and electron pockets at X/M stabilizes an $s_\pm$-wave superconducting state [**74-76**]. The introduction of disorder through Te and Al doping suggests a controlled tuning of inter-pocket scattering, which could enhance electron pairing and potentially lead to superconducting states, as previously reported in related materials [**71,77-79**].

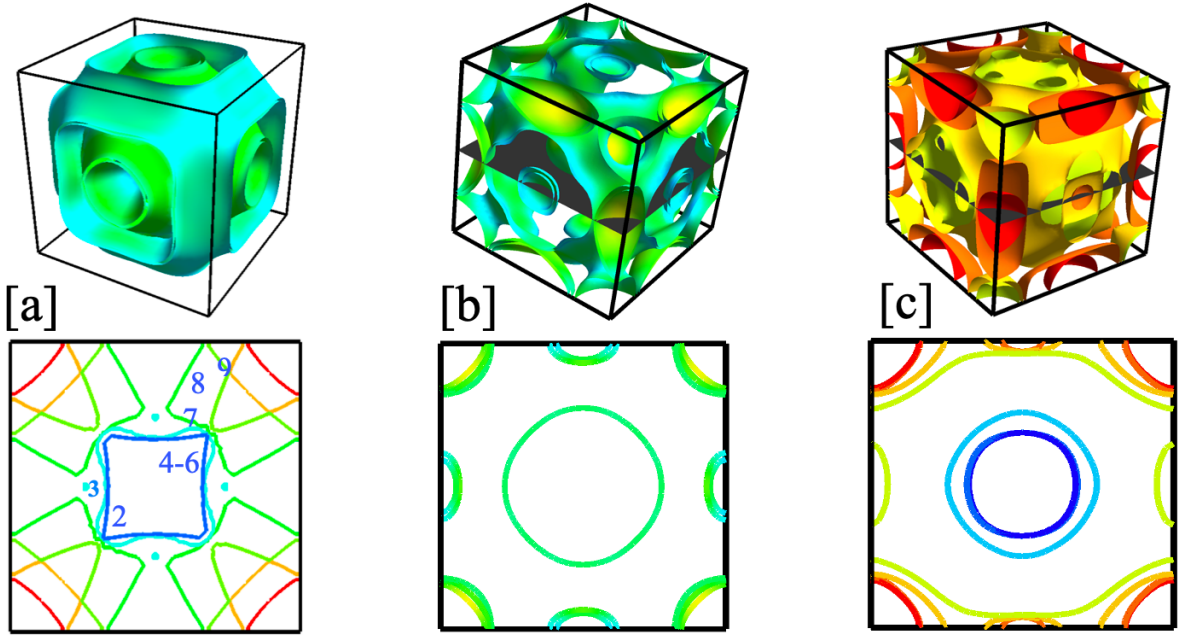

**Figure 8.** 3D and 2D projected Fermi surfaces (with SOC) of (a) YbSb, (b) Yb(Sb$_{0.97}$Te$_{0.03}$), and (c) Yb(Sb$_{0.93}$Te$_{0.035}$Al$_{0.035}$) in the rock-salt crystal structure. The Fermi surface of YbSb exhibits well-defined electron pockets with high connectivity.

## 4. Conclusion

Our study employs DFT to demonstrate that disorder is a powerful means of tuning the electronic structure and Fermi surface topology of Yb-based monopnictides. Using cubic YbSb as a model system, we introduce site-selective disorder at the Sb site via Te and Al doping, guided by their thermodynamic stability and alloying behavior. Te substitution introduces electron-like states at the X and L points by lowering the conduction band minimum by approximately 0.2 eV, while Al doping suppresses hole-like states at Γ by shifting the valence band maximum downward by about 0.15 eV. These doping-induced modifications drive a transition from a semimetallic to a narrow-gap semiconducting state, paralleling electronic trends observed in related systems such as LaBi and LaSb.

Convex hull analysis indicates that moderate doping (3% or less) with Te is feasible in YbSb, while excessive substitution leads to phase segregation. Our work shows that mechanochemical treatment of pure Yb and Sb, followed by annealing, produces a pure YbSb phase and suggests that this technique might be optimized further for incorporating Te and Al. The evolution of the Fermi surface reveals a disorder-driven transition from an interconnected electron–hole pocket structure in pristine YbSb to a more fragmented topology in doped variants, with Te doping enlarging electron pockets at the X and L points and Al doping reducing hole pockets at Γ, thereby effectively suppressing inter-pocket scattering.

Given the critical role of inter-pocket scattering in quantum transport phenomena—including superconductivity and extreme magnetoresistance—this tunability presents promising opportunities for optimizing functional properties. By linking disorder-induced band structure evolution to Fermi surface modifications, our findings establish disorder as a strategic parameter for engineering the electronic topology of rare-earth monopnictides and provide a framework for tailoring electronic properties in quantum materials. These insights lay the foundation for further theoretical and experimental investigations, particularly in designing rare-earth monopnictides for superconducting, thermoelectric, and topological applications.

## 5. Acknowledgements

The work by T.D.R., Y.M., and P.S. at Ames National Laboratory was supported by the Division of Materials Science and Engineering of the Office of Basic Energy Sciences, Office of Science of the U.S. Department of Energy (DOE). Ames National Laboratory is operated by Iowa State University for the U.S. DOE under Contract No. DE-AC02-07CH11358. M.E. and C.B. are grateful to the DOE for the assistantship and opportunity to participate in the SULI program. Their work at the Ames National Laboratory was supported by the U.S. Department of Energy Office of Science, Science Undergraduate Laboratory Internships (SULI) program.## 6. References